# Mapping the Excitation Mechanisms in the LINER I Active Galactic Nucleus NGC 5005: Positive Feedback and a Thin LINER Cocoon

Anna Trindade Falcão[1], G. Fabbiano[1], M. Elvis[1], P. Zhu[1], S. Kraemer[2], W. P. Maksym[3], R. Middei[1,4,5], and D. L. Król[1]
[1] Harvard-Smithsonian Center for Astrophysics, 60 Garden St., Cambridge, MA 02138, USA; annatrindadefalcao@gmail.com
[2] Institute for Astrophysics and Computational Sciences and Department of Physics, The Catholic University of America, Washington, DC 20064, USA
[3] NASA Marshall Space Flight Center, Huntsville, AL 35812, USA
[4] INAF Osservatorio Astronomico di Roma, Via Frascati 33, 00078 Monte Porzio Catone, RM, Italy
[5] Space Science Data Center, Agenzia Spaziale Italiana, Via del Politecnico snc, 00133 Roma, Italy

## Abstract

We present a spatially resolved Baldwin–Phillips–Terlevich analysis of the narrow-line region (NLR) in the low-ionization nuclear emission-line region (LINER) I galaxy NGC 5005 using Hubble Space Telescope narrowband imaging of [O III]$\lambda$5007, H$\beta$, H$\alpha$, and [S II]$\lambda\lambda$6717,6731. With a resolution of $\lesssim 0''.1$ ($\lesssim 10$ pc at $z = 0.003$), we dissect the NLR into H II (star-forming), Seyfert, and LINERs across spatial scales extending up to $r \sim 8$ kpc from the nucleus. Our results reveal a compact nuclear region exhibiting Seyfert-like emission, consistent with photoionization by a low-luminosity active galactic nucleus (AGN). Surrounding this Seyfert-like nucleus is a thin ($\sim 20$ pc thick) higher-excitation LINER-like cocoon, likely arising from shock-excited gas in the interstellar medium (ISM). Beyond this cocoon, a centrally localized extended ($r \sim 1$ kpc) LINER-like region surrounds the Seyfert-like nucleus and cocoon, likely ionized by the AGN, while a more extended ($r \gtrsim 2$ kpc) LINER-like zone may be ionized by a combination of post-AGB stars and shocks from gas inflows. We also detect H II–like regions at both small and large scales. In the inner 500 pc, these regions may be triggered by jet–ISM interactions, potentially inducing localized star formation. At $r \sim 4$ kpc, we identify an outer H II–like region tracing a large-scale star-forming ring, where ionization is dominated by young stars.

*Unified Astronomy Thesaurus concepts:* LINER galaxies (925)

## 1. Introduction

Active galactic nuclei (AGN) play a central role in the coevolution of galaxies and their central supermassive black holes (SMBHs). The accretion of material onto the SMBH releases immense energy in the form of radiation and outflows, which can both stimulate and suppress star formation in the host galaxy, thereby regulating the growth of both the SMBH and its host (J. Silk & M. J. Rees 1998; A. C. Fabian 2012; T. M. Heckman & P. N. Best 2014). Understanding the complex interplay between these feedback mechanisms is essential for constructing a complete picture of galaxy evolution. However, the spatial resolution required to probe key components of these processes, such as the narrow-line region (NLR), a key tracer of AGN feedback, remains limited in distant galaxies. High angular-resolution studies of nearby galaxies are therefore critical to disentangle the various processes at play (e.g., T. Storchi-Bergmann et al. 2010; J. Wang et al. 2011a, 2011b; A. Paggi et al. 2012; F. K. B. Barbosa et al. 2014).

Emission-line maps, combined with the diagnostic power of Baldwin–Phillips–Terlevich (BPT) diagrams, are invaluable for studying ionization mechanisms within galaxies. BPT diagrams employ emission-line ratios of strong optical lines, such as [N II]/H$\alpha$ (N-BPT), [S II]/H$\alpha$ (S-BPT), and [O I]/H$\alpha$ (O-BPT) versus [O III]/H$\beta$, to classify regions as AGN-dominated, star-forming, or exhibiting low-ionization nuclear emission-line region (LINER) characteristics (e.g., J. A. Baldwin et al. 1981; S. Veilleux & D. E. Osterbrock 1987; L. J. Kewley et al. 2006). These diagrams enable spatially resolved identification of coexisting excitation mechanisms within a single galaxy.

Initially identified by T. M. Heckman (1980), LINERs are characterized by their distinctive optical line ratios (e.g., L. J. Kewley et al. 2006) and can occur over a wide range of spatial scales, from nuclear ($\lesssim 10$ pc) regions to kiloparsec-scale structures (e.g., I.-T. Ho et al. 2014; F. Belfiore et al. 2016). A variety of mechanisms have been proposed to explain LINER-like emission, including shock excitation (e.g., T. M. Heckman 1980; M. A. Dopita & R. S. Sutherland 1995; I.-T. Ho et al. 2014), low-luminosity AGN activity (e.g., J. P. Halpern & J. E. Steiner 1983; T. Storchi-Bergmann et al. 1997; L. C. Ho & C. Y. Peng 2001; J. S. Ulvestad & L. C. Ho 2001), and postasymptotic giant branch stars (e.g., L. Binette et al. 1994; C. Kehrig et al. 2012; F. Belfiore et al. 2016).

Advances in integral field unit (IFU) spectroscopy (e.g., Very Large Telescope/MUSE, G. Cresci et al. 2015; AAOmega/SPIRAL, R. Sharp & J. Bland-Hawthorn 2010) and large surveys (e.g., Sloan Digital Sky Survey-IV/MaNGA, F. Belfiore et al. 2015; K. Bundy et al. 2015; S7, M. A. Dopita et al. 2015; R. L. Davies et al. 2016, 2017; L. S. Pilyugin et al. 2020) have revealed the complex, multiphase nature of the interstellar medium (ISM), with Seyfert-like, LINER-like, and star-forming regions often coexisting in close proximity. Nonetheless, the unmatched spatial resolution of the Hubble Space Telescope (HST) ($\lesssim 0''.1$) remains essential for resolving the innermost structures.

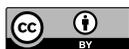

W. P. Maksym et al. (2016) conducted a pilot HST BPT study of the nearby Compton-thick Seyfert 2 galaxy NGC 3393, uncovering a predominantly Seyfert-like NLR surrounded by a previously unknown LINER-like cocoon. This was the first such structure clearly revealed in high-resolution optical imaging studies. A similar structure was identified earlier in NGC 5643 by G. Cresci et al. (2015) using ground-based data, where the nucleus and ionization cones exhibit Seyfert-like excitation, enveloped by a LINER-like cocoon.

J. Ma et al. (2021) extended this work by analyzing a sample of seven nearby type 2 Seyfert galaxies. Their results showed a consistent picture in which the nuclear regions and ionization cones exhibit Seyfert-like emission, likely powered by AGN photoionization, while being embedded within LINER-like cocoons.

However, LINER-dominated galaxies have not yet been the focus of similar studies. To address this gap, we obtained narrowband HST imaging of NGC 5005.

NGC 5005 is a nearby SAB(rs) galaxy at a distance of $D \sim 17$ Mpc ($1'' \sim 100$ pc, from NASA/IPAC Extragalactic Database),[6] making it an ideal laboratory for resolving the distribution of different ionization phases in a LINER-dominated galaxy. This galaxy is known to host a LINER at its center (L. C. Ho et al. 1997), and has been classified both as a LINER I and a low-luminosity AGN (e.g., M. L. Saade et al. 2022). Moreover, evidence for a broad H$\alpha$ component has been reported (J. R. Stauffer 1982; L. C. Ho et al. 1995), as well as a possible Seyfert 2 nucleus (R. Maiolino & G. H. Rieke 1995).

Morphologically, NGC 5005 shows evidence of a weak, round bar roughly aligned with the galaxy's major axis (position angle (P.A.) = 65°; G. Vaucouleurs et al. 1991; K. Sakamoto et al. 2000). Although weak, this bar may still influence gas dynamics by driving shocks in the central molecular gas reservoir (M. Das et al. 2003). Imaging studies (e.g., R. W. Pogge et al. 2000) reveal a dust-obscured nucleus and clumpy optical line emission within $r \sim 100$ pc, surrounded by filamentary and diffuse emission extending out to $r \sim 300$ pc.

Molecular gas studies by K. Sakamoto et al. (2000) identified a CO-emitting circumnuclear ring at $r \sim 3$ kpc, connected via inflowing gas streams to a nuclear disk at $r \sim 1$ kpc. In the radio, R. D. Baldi et al. (2018) reported a $\sim 2$ kpc jet aligned along P.A. = 27°.

In X-rays, early ASCA observations suggested a highly obscured nucleus, with G. Risaliti et al. (1999) reporting a column density of $N_H > 10^{24}$ cm$^{-2}$. However, later observations with Chandra and XMM-Newton yielded a significantly lower column density of $N_H \sim 1.5 \times 10^{20}$ cm$^{-2}$ (M. Guainazzi et al. 2005), a result subsequently confirmed by M. Brightman & K. Nandra (2011) and G. Younes et al. (2012), suggesting that the nucleus had been previously misclassified as Compton-thick.

To explore whether LINER-dominated galaxies also exhibit LINER-like cocoons surrounding the central nuclear region, we obtained 11 orbits of HST/Wide Field Camera 3 (WFC3) as part of a joint Chandra/HST program targeting NGC 5005. In this paper, we present the results of the HST observations and analyze the inner $\sim 8$ kpc using methods similar to those employed by W. P. Maksym et al. (2016) and J. Ma et al. (2021).

**Table 1**
HST/WFC3 Observations of NGC 5005 Used in This Work

| ObsID | Date | Exp. Time (s) | Filter | Band (rest) |
|---|---|---|---|---|
| ieqt02010 | 2022-06-26 | 5928 | F487N | H$\beta$ |
| ieqt02020 | 2022-06-26 | 4728 | ⋯ | ⋯ |
| ieqt01010 | 2022-06-25 | 5928 | F502N | [O III] |
| ieqt01020 | 2022-06-25 | 4728 | ⋯ | ⋯ |
| ieqt04020 | 2022-05-13 | 695 | F547M | blue continuum |
| ieqt04010 | 2022-05-13 | 695 | F658N | H$\alpha$+[N II] |
| ieqt03010 | 2022-06-16 | 5333 | F673N | [S II] |
| ieqt04030 | 2022-05-13 | 695 | F814W | red continuum |

## 2. Observations and Data Reduction

### 2.1. Observations

NGC 5005 was observed as part of a joint Chandra/HST program (Program ID: 16837, PI: Fabbiano) aimed at constructing high-resolution, multiwavelength maps of the nuclear regions of nearby active galaxies. The HST observations employed WFC3 with narrowband filters to image key diagnostic lines, including H$\beta$, [O III], H$\alpha$, and [S II]. Table 1 summarizes the details of these observations, including obsids, dates, exposure times, and filters used. All HST data used in this study can be found in MAST doi:10.17909/cjbp-dj36.

At the redshift of NGC 5005 ($z = 0.003156$), the selected narrowband filters—F487N, F502N, F658N, and F673N—cover H$\beta$, [O III], H$\alpha$+[N II], and [S II], respectively. Continuum emission was sampled using the F547M and F814W filters for the blue and red bands, respectively, to enable accurate subtraction of the stellar continuum from the narrowband images. The WFC3 observations achieved an angular resolution of $\sim 0.''04$ per pixel, corresponding to a physical scale of $\sim 4$ pc at $D \sim 17$ Mpc. These high-resolution images provide an unprecedented view of the nuclear region, enabling a detailed analysis of the spatial distribution of ionized gas. For the WFC3/UVIS filters listed in Table 1, the empirical point-spread function has a full width at half-maximum (FWHM) $\lesssim 0.''1$.[7]

### 2.2. Data Reduction

Data reduction was performed using the `DrizzlePac`[8] software package (S. L. Hoffmann et al. 2021), following standard HST pipeline procedures. Cosmic rays were removed using `Astrodrizzle` for filters with multiple exposures. For single-exposure filters (F547M, F658N, and F814W), cosmic rays were manually removed using the `PyRAF` tool *imedit*. Subexposure astrometry was aligned using the `TweakReg` and `TweakBack` tools, with all final images aligned to the F502N filter, which had the longest total exposure time.

The images were drizzled to a final pixel scale of $0.''04$, and background subtraction was performed using median values from emission-free regions. Flux calibration was applied using

---

[6] https://ned.ipac.caltech.edu/
[7] https://hst-docs.stsci.edu/wfc3ihb/chapter-6-uvis-imaging-with-wfc3/6-6-uvis-optical-performance#id-6.6UVISOpticalPerformance-6.6.16.6.1PSFWidthandSharpness
[8] https://www.stsci.edu/scientific-community/software/drizzlepac

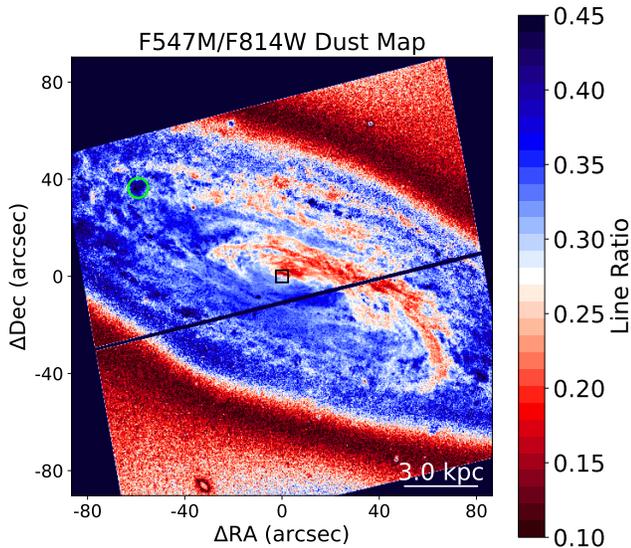

**Figure 1.** Color map derived from the image ratio F547M/F814W (blue/red continua). North is up and east is to the left. This ratio map shows clearly the nuclear dust lane (in dark red). The color scale is chosen to enhance similar features. The green circle on the top left represents the low-dust region used in our analysis (see the Appendix). The black square has dimensions of $5'' \times 5''$, and marks the location of the nuclear region.

the PHOTFLAM and PHOTBW header keywords, converting the image units to erg s$^{-1}$ cm$^{-2}$.

### 2.3. Narrow-line Imaging

Dust presents a significant challenge in our analysis, as illustrated by the F547M/F814W (blue continuum/red continuum) ratio map in Figure 1. Regions most affected by dust extinction appear in dark red, where continuum subtraction becomes less reliable. To mitigate this, we followed the approach of W. P. Maksym et al. (2021), as detailed in the Appendix, which assumes that the color of a low-dust region (indicated by the green circle in Figure 1) represents the intrinsic stellar population of the galaxy. Using this reference, we applied a wavelength-dependent reddening correction to rescale the emission in each narrowband filter prior to continuum subtraction, following the method of W. P. Maksym et al. (2021).

Due to the relatively broad bandwidth of the F658N filter, the H$\alpha$ emission is significantly contaminated by [N II]$\lambda\lambda$6548, 6584 and must be corrected. Existing IFU data for NGC 5005 reveal spatial variations in local H$\alpha$/[N II] ratios across the NLR (E. E. Richards et al. 2015). To account for uncertainties in the [N II] contribution, we compared the total continuum-subtracted H$\alpha$+[N II]$\lambda\lambda$6548, 6584 flux in our image with the total H$\alpha$ image flux reported by E. E. Richards et al. (2015) over the same region, finding an average [N II] contribution of $\sim$55% across the full field of view. In the central $5''$ (500 pc), comparison with the H$\alpha$ flux from L. C. Ho et al. (1997) indicates a higher [N II] contribution of $\sim$70%. Based on these comparisons, we adopt a 45% H$\alpha$ contribution to the total emission in the F658N bandpass throughout this paper, and use 30% as a conservative lower limit.

## 3. Results

### 3.1. Narrow-line Region Morphology

The continuum-subtracted emission-line images of NGC 5005 reveal an elongated morphology along the northeast–southwest direction (Figure 2). We adopt the [O III] continuum centroid (hereafter, optical centroid) as the reference nucleus location, following, e.g., T. C. Fischer et al. (2018). The $80'' \times 80''$ (8 kpc$^2$; top panels of Figure 2) narrowband images of [O III], H$\beta$, H$\alpha$, and [S II] show an extended ($r \sim$ 3–4 kpc) NLR aligned with the galaxy's disk. The H$\alpha$ image highlights an enhanced ring-like component at $r_{\rm maj} \sim$ 4 kpc and $r_{\rm min} \sim$ 1.2 kpc, oriented at a position angle of P.A. = 68°. Along this ring, a bright, clumpy H$\alpha$-emitting region is visible to the southeast of the nucleus (indicated by a black square), colocated with the southern radio lobe reported by R. D. Baldi et al. (2018; see Section 4), suggesting interactions between the radio jet and the surrounding NLR gas.

In the central $5'' \times 5''$ region (500 pc$^2$; bottom panels of Figure 2), the ionized gas is concentrated in compact clumps within $r \sim$ 100 pc of the nucleus (indicated by a black cross), and surrounded by fan-shaped filaments and diffuse emission extending out to $r \sim$ 300 pc. A bubble-like feature is observed in H$\alpha$ south of the nucleus, though it appears less evident, or even absent, in the [O III], H$\beta$, and [S II] images.

Note that the H$\alpha$ emission appears comparatively weaker due to the shorter exposure time of the F658N narrowband filter (Table 1), relative to the exposures of the [O III], H$\beta$, and [S II] bands.

### 3.2. S-BPT Mapping

We use the S-BPT diagram, plotting log([O III]/H$\beta$) versus log([S II]/H$\alpha$), as described by L. J. Kewley et al. (2006), to construct excitation maps for NGC 5005. Each pixel in the narrow-line images (Figure 2) is classified according to its position on the S-BPT diagram, enabling identification of regions dominated by different excitation mechanisms across the NLR.

Flux ratios are calculated on a per-pixel basis from the continuum-subtracted narrowband images and plotted in the top-left panel of Figure 3. Only pixels with a signal-to-noise ratio (S/N) greater than 3$\sigma$ are included in the analysis. The 3$\sigma$ threshold was determined from pixel count statistics in non-background-subtracted images. Data values were first converted to counts using the exposure time. A background-dominated region, free of line emission, was used to estimate noise, with the 3$\sigma$ threshold defined as 3 times the standard deviation of the pixel counts. This count threshold was then converted to flux using the exposure map and applied to the background-subtracted images to generate 3$\sigma$ emission-line masks, following the approaches of W. P. Maksym et al. (2016) and J. Ma et al. (2021).

In the S-BPT diagram, regions are color-coded to represent different excitation mechanisms: red for Seyfert-like, green for LINER-like, and yellow for H II–like (star-forming) emission. Each WFC3 pixel ($0''.04 \times 0''.04$, or 4 pc$^2$) is treated as a single data point in the diagram.

Following the methodology of W. P. Maksym et al. (2016), we generate spatially resolved S-BPT (hereafter, sr-S-BPT) maps by assigning each pixel a classification based on its diagram location. These maps reveal the spatial distribution of different excitation mechanisms within the galaxy. The top-right, bottom-left, and bottom-right panels of Figure 3 show three such maps: an $80'' \times 80''$ field (8 kpc$^2$; top-right panel, hereafter Map 1); a $5'' \times 5''$ central region (500 pc$^2$; bottom-left panel, hereafter Map 2); and a $2'' \times 2''$ nuclear zoom

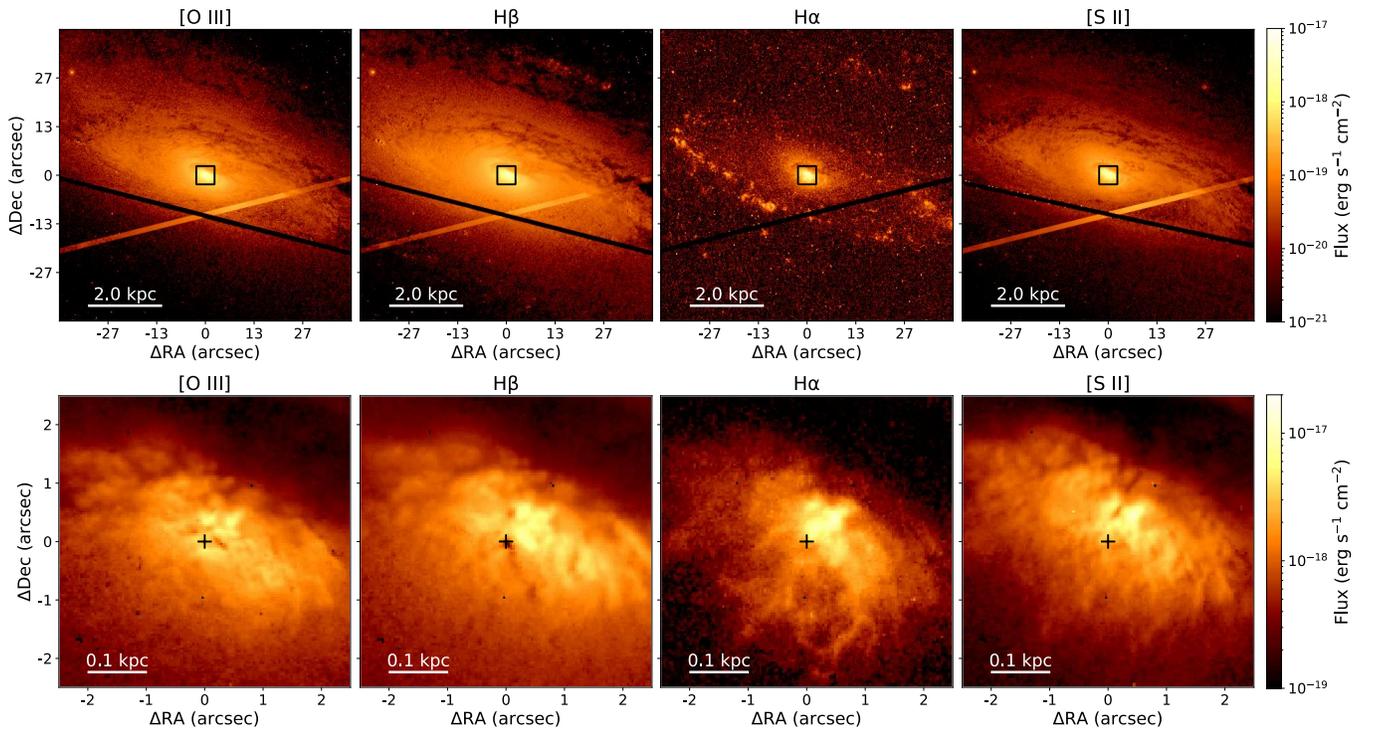

**Figure 2.** Narrow-line images of the NLR and inner regions of NGC 5005 covering prominent emission lines. The top panels show an 80″ × 80″ (8 kpc × 8 kpc) field, while the bottom panels zoom into a 5″ × 5″ (500 pc × 500 pc) central region. Stellar continuum has been subtracted in all images, and only pixels with signal-to-noise ratios greater than $3\sigma$ are shown. The black squares in the top panels indicate the area shown in the bottom panels. The black crosses in the bottom panels mark the location of the optical nucleus. All images are shown in log scale. In all panels, north is up and east is to the left.

(200 pc$^2$; bottom-right panel, hereafter Map 3). Pixels below the $3\sigma$ threshold are shown in black.

The S-BPT diagram in Figure 3 (top-left panel) reveals that in Map 1, 54.80% of pixels fall within the H II–like classification, 45.13% are LINER-like, and only 0.07% are Seyfert-like. In Maps 2 and 3, the distribution shifts significantly: H II–like pixels drop to 1.94% and 4.59%, while LINER-like regions dominate at 97.98% and 95.00%, respectively. Seyfert-like classifications increase slightly to 0.08% and 0.41%.

*Map 1.* H II–like emission forms an elliptical ring morphology, enhanced in the outer regions, closely matching the H$\alpha$ structure observed in Figure 2. LINER-like emission is embedded within this ring, aligned along the northeast–southwest axis. No significant Seyfert-like emission is detected in this map.

*Map 2.* In this zoomed-in region, Seyfert-like emission is confined to the central ∼20 pc. H II–like emission is patchy and localized within ∼200 pc from the nucleus, while LINER-like excitation dominates the map.

*Map 3.* Within the inner 200 pc, LINER-like emission remains dominant, but Seyfert-like excitation is clearly detected at the optical centroid (marked by a black "X"), forming a compact "primary cone" covering 7 pixels. A secondary Seyfert-like region (hereafter, the "secondary cone") appears ∼40 pc northwest of the nucleus, spanning 3 pixels. West of the nucleus, four clumpy H II–like regions, comprising 111 pixels, are found at $r \sim 90$ pc.

The S-BPT diagram in Figure 3 shows two subpopulations of LINER-classified pixels: a horizontal grouping (highlighted with a green ellipse) and a diagonal branch (blue ellipse). The latter appears spatially separated from the main group of LINER pixels, hinting at structurally different components. To explore this, we reclassified LINER pixels by log([O III]/H$\beta$): those with log([O III]/H$\beta$) ⩾ 0.28 were assigned to a "blue LINER" group, and the rest to a "green LINER" group. These reclassified groups and their spatial distributions are shown in Figure 4.

The new sr-S-BPT maps in Figure 4 confirm that the two LINER groups are associated with different structures in the galaxy. Green LINER-like pixels dominate the emission in the innermost regions, while blue LINER-like pixels form a surrounding structure around the Seyfert-like "cones," resembling the LINER-like cocoon morphology reported by W. P. Maksym et al. (2016) and J. Ma et al. (2021).

In the reclassified S-BPT diagram (Figure 4, top-left panel), pixel fractions remain similar to the previous classification for Map 1: 54.80% H II–like, 45.08% LINER-like, 0.07% Seyfert-like, and 0.05% cocoon-like. In Maps 2 and 3, H II–like pixels represent 1.94% and 4.59%, while LINER-like fractions decrease to 97.44% and 92.31%. Seyfert-like pixels account for 0.08% and 0.41%, and cocoon-like pixels for 0.54% and 2.69%, respectively.

When assuming a lower limit for H$\alpha$ emission in the F658N filter (see Section 2.3), the contributions shift: Map 1 contains 48.74% H II–like pixels, 51.06% LINER-like, 0.12% Seyfert-like, and 0.09% cocoon-like. In Maps 2 and 3, the H II–like fraction drops to 0.36% and 1.12%, while LINER-like fractions increase to 98.62% and 94.85%. Seyfert- and cocoon-like regions contribute 0.03% and 0.11%, and 1.00% and 3.92%, respectively. These shifts occur because in star-forming regions, Balmer lines such as H$\alpha$ are typically stronger than forbidden lines such as [S II] (e.g., D. E. Osterbrock & G. J. Ferland 2006). A lower assumed H$\alpha$ emission increases the [S II]/H$\alpha$ ratios, shifting points rightward on the diagram and into the LINER regime.

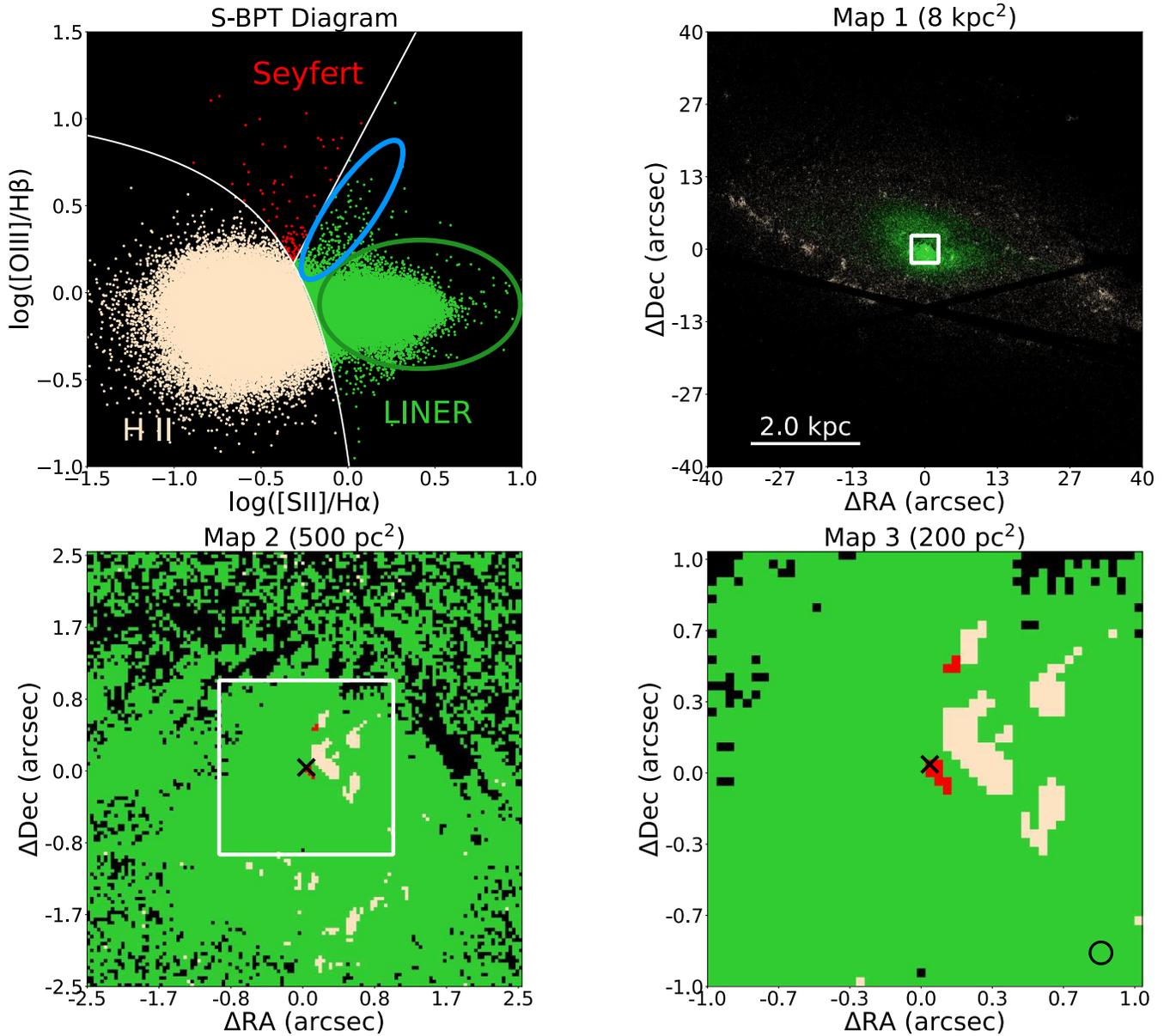

**Figure 3.** Top-left panel: S-BPT diagram of the inner $80'' \times 80''$ region in NGC 5005. Each $0\rlap{.}''04 \times 0\rlap{.}''04$ pixel is plotted as a single data point. Classification criteria from L. J. Kewley et al. (2006) are indicated as solid white lines. Red, green, and yellow points correspond to Seyfert-like, LINER-like, and H II–like (star-forming) excitation, respectively. A $3\sigma$ S/N cut has been applied. The green and blue ellipses highlight the two distinct LINER subpopulations, horizontal and diagonal groups, respectively (see the main text for details). Top-right panel: spatially resolved S-BPT (sr-S-BPT) map of the same $80'' \times 80''$ region, constructed from continuum-subtracted narrow-line images. Each HST pixel is color-coded based on its classification in the top-left diagram. Black pixels indicate regions below the $3\sigma$ detection threshold. The white square outlines the central region shown in the bottom-left panel. Bottom-left panel: zoom-in of the inner $5'' \times 5''$ (500 pc × 500 pc) region. The white square marks the area shown in the bottom-right panel. Bottom-right panel: further zoom-in of the inner $2'' \times 2''$ (200 pc × 200 pc) region. The black "X" marks the [O III] continuum peak, interpreted as the position of the nucleus. The black circle on the bottom-right indicates the WFC3/UVIS PSF FWHM. In all spatially resolved maps, north is up and east is to the left.

## 4. Discussion

The S-BPT diagram and corresponding sr-S-BPT maps in Figure 4 provide insights into the galaxy's morphology and the various excitation mechanisms at play. In Section 4.1, we focus on structures within the inner 500 pc, including Seyfert-like and cocoon-like emission ($r < 100$ pc; Section 4.1.1) and the clumpy H II–like regions at $r \sim 500$ pc (Section 4.1.2). Section 4.2 addresses features observed at larger radii, including the central LINER-like emission at $r \sim 1$ kpc (Section 4.2.1), extended LINER-like emission beyond $r \gtrsim 2$ kpc (Section 4.2.2), and the H II–like emission in a ring-like structure at $r \sim 4$ kpc (Section 4.2.3).

### 4.1. Resolved Structures in the Inner 500 pc

#### 4.1.1. Seyfert-like Nucleus and Surrounding LINER-like Cocoon

The sr-S-BPT maps in Figure 4 reveal that within the central 100 pc (bottom-right panel, Map 3), Seyfert-like emission is confined to the innermost region, extending over $\sim 10$–20 pc. These Seyfert-classified pixels exhibit a biconical morphology

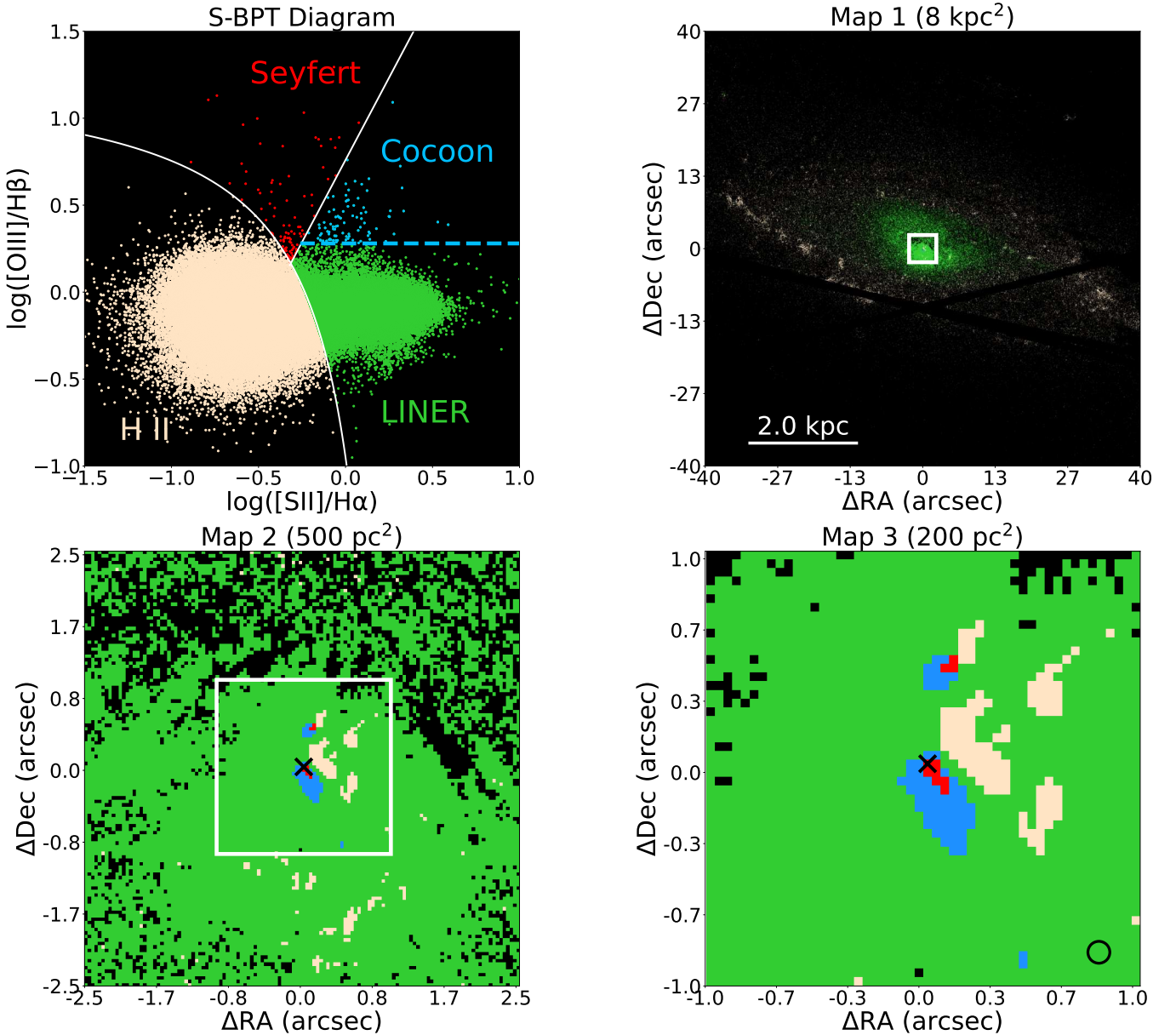

**Figure 4.** Same as Figure 3, but with the LINER-like pixels separated into two groups based on their log([O III]/Hβ) values. Pixels with log([O III]/Hβ) ⩾ 0.28 are shown in blue, and those with lower log([O III]/Hβ) values in green. The blue dashed line in the top-left panel marks the division between the two LINER groups. In all spatial maps, north is up and east is to the left. The black circle in the bottom-right panel indicates the WFC3/UVIS PSF FWHM.

(10 pixels in total; see also Section 4.1.2), with each cone enclosed by a thin LINER-like cocoon ∼10–20 pc in thickness.

The primary Seyfert-like cone appears colocated with the position of the optical centroid (marked by a black X), suggesting that the emission is primarily driven by AGN photoionization. The presence of an AGN at this location is further supported by the detection of a hard X-ray source in Chandra/ACIS-S observations (A. Trindade Falcão et al. 2025, in preparation). However, we note that fast shocks ($v > 500$ km s$^{-1}$) with photoionizing precursors can also produce Seyfert-like ratios in BPT diagrams (e.g., M. G. Allen et al. 2008), and cannot be ruled out as a contributing mechanism (see Section 4.1.2).

The surrounding LINER-like cocoon is consistent with similar structures observed in Seyfert-dominated galaxies (e.g., G. Cresci et al. 2015; W. P. Maksym et al. 2016; J. Ma et al. 2021) and likely arises from shock excitation. For instance, outflows propagating from the nucleus at $v \sim 200$ km s$^{-1}$ may interact with and shock the surrounding ISM, producing the observed cocoon-like emission (see Section 4.1.2). Alternatively, this emission may be the result of "filtered" AGN radiation. In this scenario, regions outside the ionization cones receive only attenuated ionizing flux due to obscuration by a torus or hollow biconical outflow (S. B. Kraemer et al. 2008; T. M. Heckman & P. N. Best 2014), leading to a stratified ionization structure.

To quantify the AGN's ionizing power, we assume that the [O III] emission across the NLR is produced entirely by AGN photoionization. We measure the extinction-corrected [O III] λ5007 imaging flux out to $r = 1.5$ kpc from the nucleus, obtaining $F_{\text{[O III]}} = 3.4 \times 10^{-14}$ erg s$^{-1}$ cm$^{-2}$. Assuming a distance of $D = 17$ Mpc, the corresponding integrated [O III] luminosity is $L_{\rm [O\,III]} = 1.2 \times 10^{39}\,{\rm erg\,s^{-1}}$.

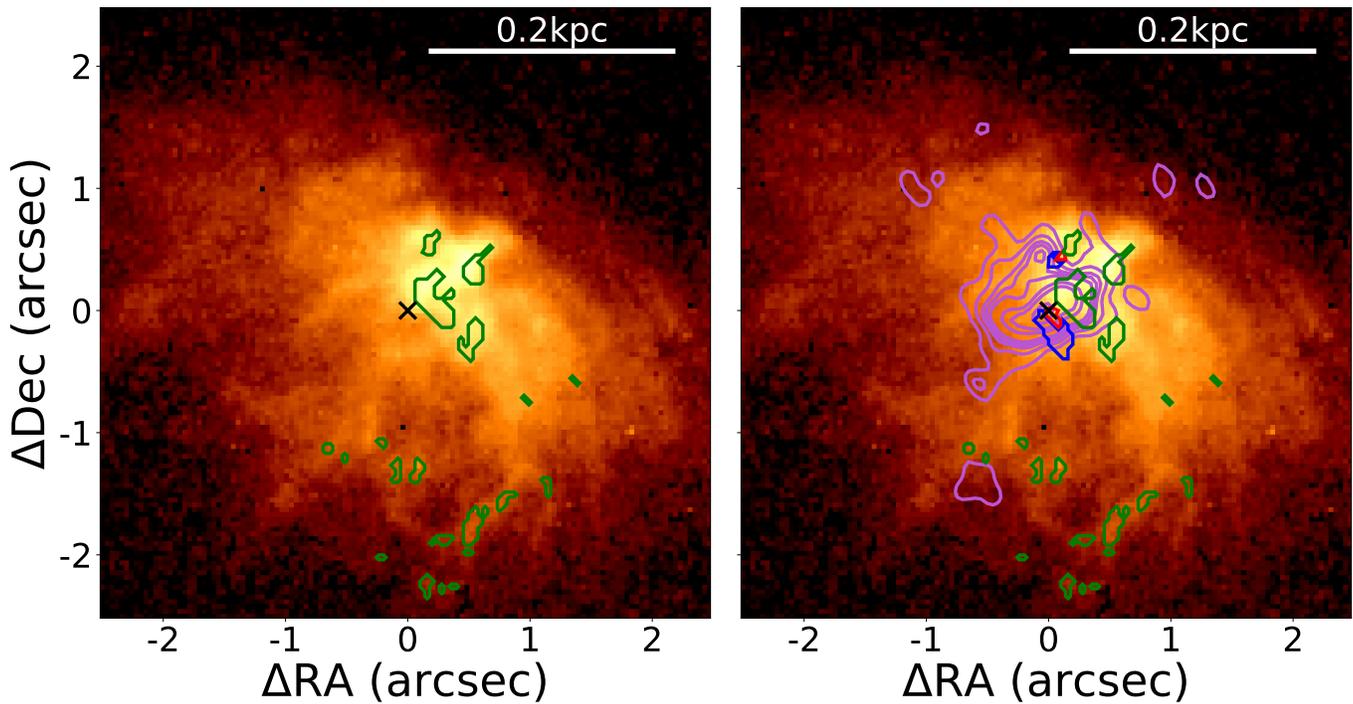

**Figure 5.** Left panel: H$\alpha$ narrow-line image of NGC 5005, as in Figure 2. Green contours highlight H II–like emission identified in the sr-S-BPT maps within the inner 500 pc. Right panel: same H$\alpha$ image, overlaid with full-resolution 1.5 GHz eMERLIN radio contours (in purple). Red, blue, and green contours correspond to Seyfert-like, cocoon-like, and H II–like emission, respectively, as classified in the sr-S-BPT maps. In both panels, north is up and east is to the left.

There is ongoing debate about the appropriate bolometric correction factor for converting $L_{\rm [O\,III]}$ to $L_{\rm bol}$, due to the role of internal extinction and potential luminosity dependence (e.g., T. M. Heckman et al. 2005; A. Lamastra et al. 2009; H. Netzer 2009). One approach is to use the empirical correlation between [O III]$\lambda$5007 luminosity and 3–20 keV X-ray luminosity from T. M. Heckman et al. (2005). Adopting a conversion factor of $L_{3-20\,{\rm keV}}/L_{2-10\,{\rm keV}} = 1.17$, estimated using PIMMS,[9] we infer $L_{2-10\,{\rm keV}} = 1.4 \times 10^{41}\,{\rm erg\,s^{-1}}$. Applying the bolometric correction from F. Duras et al. (2020), we estimate $L_{\rm bol} = 2.1 \times 10^{42}\,{\rm erg\,s^{-1}}$. It is important to note, however, that this relation is based on the observed (i.e., uncorrected for extinction) [O III] luminosity.

An alternative method is to use the correlation between $L_{2-10\,{\rm keV}}$ and the extinction-corrected $L_{\rm [O\,III]}$ from A. Lamastra et al. (2009), which yields $L_{2-10\,{\rm keV}} = 1.3 \times 10^{40}\,{\rm erg\,s^{-1}}$. Applying their associated bolometric correction, we obtain $L_{\rm bol} = 1.0 \times 10^{41}\,{\rm erg\,s^{-1}}$. In this case, however, the correlation is based on observed X-ray luminosities, rather than intrinsic ones.

Given the assumptions and uncertainties in both methods, we interpret these as upper and lower limits on the AGN's bolometric luminosity, respectively. This places the AGN in NGC 5005 in the low-luminosity regime, with important implications for understanding the origin of the central LINER-like emission discussed in Section 4.2.1. A more precise determination of the X-ray and bolometric luminosities will be presented in a forthcoming study (A. Trindade Falcão et al. 2025, in preparation), which will analyze the Chandra X-ray data in detail.

_________
[9] https://cxc.harvard.edu/toolkit/pimms.jsp

*4.1.2. Clumpy H II–like Region*

Within the inner 500 pc, patchy regions of H II–like excitation are observed in the sr-S-BPT maps (234 H II–like pixels in Map 2, Figure 4).

Figure 5 (left panel) compares the spatial distribution of these H II–like pixels (green contours) with the morphology of the H$\alpha$ narrow image. The H II–like regions are colocated with two key features in the H$\alpha$ image:

1. A peanut-shaped structure (70 pc × 30 pc), northwest of the optical centroid, flanked by two regions of enhanced line emission.
2. A bubble-like structure (190 pc × 130 pc), south of the nucleus.

The right panel of Figure 5 shows the H$\alpha$ morphology overlaid with the high-resolution ($\lesssim 0\farcs 2$) 1.5 GHz nuclear radio jet observed by R. D. Baldi et al. (2018) (purple contours), along with contours highlighting H II–like (green), Seyfert-like (red), and cocoon-like (blue) regions. Several radio features overlap with regions of distinct excitation, suggesting a connection between nuclear jet activity and excitation of the ionized gas. Specifically:

1. A radio knot north of the optical centroid is roughly colocated with Seyfert-like (red) and cocoon-like (blue) pixels at the location of the secondary cone (see Section 4.1.1). Since Seyfert-like BPT ratios can also result from fast shocks with photoionizing precursors (e.g., M. G. Allen et al. 2008), it is possible that the jet is shocking the ISM in this region, producing the observed Seyfert-like ratios. Supporting this interpretation, when adopting a lower H$\alpha$ contribution in the F658N filter bandpass, Seyfert-like pixels in the secondary cone transition to a cocoon-like classification, likely tracing shock-excited gas in the ISM.

2. Northwest of the optical centroid, the jet appears colocated with clumpy H II–like regions (green), suggesting that interactions with the ISM may be triggering localized star formation. This aligns with the idea of jet-induced positive feedback, wherein compression of the ISM by the radio jet promotes star formation.

3. South of the optical centroid, H II–like contours trace the bubble-like feature seen in Hα, oriented nearly perpendicular to the radio jet, with a relative inclination of ∼70°. Similar plumes of ablated gas carried by hot bubbles have been observed to propagate perpendicular to galactic disks in nearby Seyfert galaxies and are thought to result from jet–ISM interactions (e.g., D. Mukherjee et al. 2018; W. P. Maksym et al. 2021; G. Fabbiano et al. 2022). The Hα bubble may therefore trace partially ionized gas surrounding such a structure. Hydrodynamic simulations by D. Mukherjee et al. (2018) show that inclined jets impacting the galactic disk can shock and ablate clouds, forming arc-like structures. As the jet decelerates, it inflates a slower, hot gas bubble that rises perpendicular to the disk, resembling the low-density Hα bubble seen in NGC 5005. This expanding hot bubble may compress the surrounding ISM, thereby triggering star formation.

An alternative interpretation is that the Hα bubble represents a wind-driven outflow, similar to the superbubble in NGC 3079 (e.g., G. Cecil et al. 2001), though on a smaller scale (by a factor of ∼5). Long-slit optical spectroscopy of NGC 5005 reveals nuclear Hα outflows with velocities of $v \sim 150\,\mathrm{km\,s^{-1}}$ and FWHM $\sim 2150\,\mathrm{km\,s^{-1}}$ (S. Cazzoli et al. 2018), consistent with this scenario.

Figure 2 reveals filamentary structures in the [S II], Hβ, and [O III] maps, suggesting a complex, multiphase ISM with several "bubbly" features. Among these, the Hα bubble might stand out as the most prominent and morphologically distinct. As wind-driven outflows expand, the gas is organized into a sequence of shocks and ionization fronts, as predicted by simulations (A. V. R. Schiano 1985, 1986; D. K. Strickland et al. 2000). In a mature bubble, this sequence typically includes a photoionized precursor, a dense shell of compressed ISM forming a standing "ring shock," the contact discontinuity between the shocked wind and ISM, a hot shocked wind region, the wind shock, and the free-flowing wind near the energy injection site. In this scenario, the wind's interaction with the ISM may be triggering star formation, observed as H II–like regions tracing the contours of the Hα bubble in Figure 5.

In NGC 3079, the $r = 1.1$ kpc superbubble exhibits unusually high [N II]λ6584/Hα ratios, indicative of strong shock excitation (S. Veilleux et al. 1994). Similarly high [N II]/Hα ratios have been reported in NGC 5005 (e.g., L. C. Ho et al. 1997). A forthcoming study (A. Trindade Falcão et al. 2025, in preparation) will present a detailed spatial and spectral analysis of this feature to further investigate the physical conditions in the bubble and the processes driving its expansion.

### 4.2. Large-scale Structures and Emission Signatures

#### 4.2.1. Central LINER Emission

Maps 1 and 2 in Figure 4 reveal a centrally localized region of LINER-like excitation within the NLR of NGC 5005. This region exhibits an elliptical morphology with dimensions of $r_{\mathrm{maj}} \sim 1.4$ kpc and $r_{\mathrm{min}} \sim 800$ pc. The central LINER-like zone encompasses the Seyfert-like nucleus and surrounding cocoon-like region (Section 4.1.1), and is itself enclosed by a more extended, diffuse LINER-like structure (see Section 4.2.2) and an outer H II–like ring (see Section 4.2.3).

This structure is reminiscent of the central low-ionization emission-line region or "cLIER" galaxy morphology described by F. Belfiore et al. (2016), characterized by extended central LINER-like emission and ongoing star formation at larger radii (see their Figure 5(c)). In NGC 5005, the central LINER-like region is likely powered by photoionization from the low-luminosity AGN (Section 4.1.1), although contributions from shocks are also plausible (Section 4.1.2). Evidence for shock excitation is supported by high [Fe II]/[P II] and [Fe II]/Paβ ratios observed in the NLR (K. Terao et al. 2016, see also Section 4.1.2).

The role of a low-luminosity AGN as the dominant ionizing source in this region will be investigated further in a forthcoming study (A. Trindade Falcão et al. 2025, in preparation), which will use X-ray observations to constrain the AGN's spectral energy distribution, as well as its X-ray and bolometric luminosities.

#### 4.2.2. Large-scale LINER Emission

Beyond $r \sim 2$ kpc, the excitation is predominantly H II–like (97.18%; see Section 4.2.3 and Figure 4, Map 1). However, 2.68% of pixels in this region exhibit LINER-like excitation, which increases to ∼15% when adopting the lower limit for Hα emission in the F658N filter bandpass (see Section 2.3). The distribution of these large-scale LINER-like pixels is shown in Figure 6 as green X's, along with Seyfert-like (0.13%, red X's) and cocoon-like (0.02%, blue X's) pixels.

In the left panel, yellow ellipses outline a region with outer dimensions of $r_{\mathrm{maj}}^{\mathrm{out}} = 4.3$ kpc and $r_{\mathrm{min}}^{\mathrm{out}} = 1.4$ kpc, and inner dimensions of $r_{\mathrm{maj}}^{\mathrm{in}} = 2.8$ kpc and $r_{\mathrm{min}}^{\mathrm{in}} = 900$ pc, which is dominated by the H II–like ring (see Section 4.2.3). LINER-like pixels are distributed along the ring, suggesting a connection. One possible origin for this extended LINER emission is ionization by postasymptotic giant branch (post-AGB) stars, which populate old stellar populations along the ring. These evolved stars, with typical temperatures of ∼30,000 K, are efficient sources of ionizing UV photons (e.g., F. Belfiore et al. 2016). In this scenario, the LINER-like emission may arise primarily from photoionization by post-AGB stars, although localized shocks could also contribute to the observed excitation in this region.

The middle panel of Figure 6 shows the 1.5 GHz low-resolution eMERLIN radio contours from R. D. Baldi et al. (2018) overlaid in orange. Most Seyfert-like and cocoon-like pixels along the ring are concentrated northwest of the nucleus. The outer radio lobes spatially coincide with the H II–like ring; although the northern lobe overlaps a region containing several Seyfert-like pixels, there is no clear evidence of enhanced excitation at the jet's point of impact.

The right panel of Figure 6 shows the CO(1-0) zeroth-moment contours from K. Sakamoto et al. (2000) overlaid in white. While some of the Seyfert-like and cocoon-like pixels in the northwestern portion of the ring may result from noise in the data, localized shocks could also contribute to their excitation. This interpretation is supported by the CO(1-0) morphology: K. Sakamoto et al. (2000) proposed the presence of a small stellar bar in NGC 5005 (see also Section 4.2.3), which drives gas into the CO ring at $r \sim 3$ kpc and into the central condensation disk ($r \sim 1$ kpc). The orientation of the bar implies that the northwest CO stream, which connects the outer CO ring to the central disk, originates from shocks along the dust lanes at the bar's leading edge. As shown in the right panel of Figure 6, Seyfert-like pixels (in red) cluster near this CO stream, reinforcing the idea that shocks contribute to their excitation.

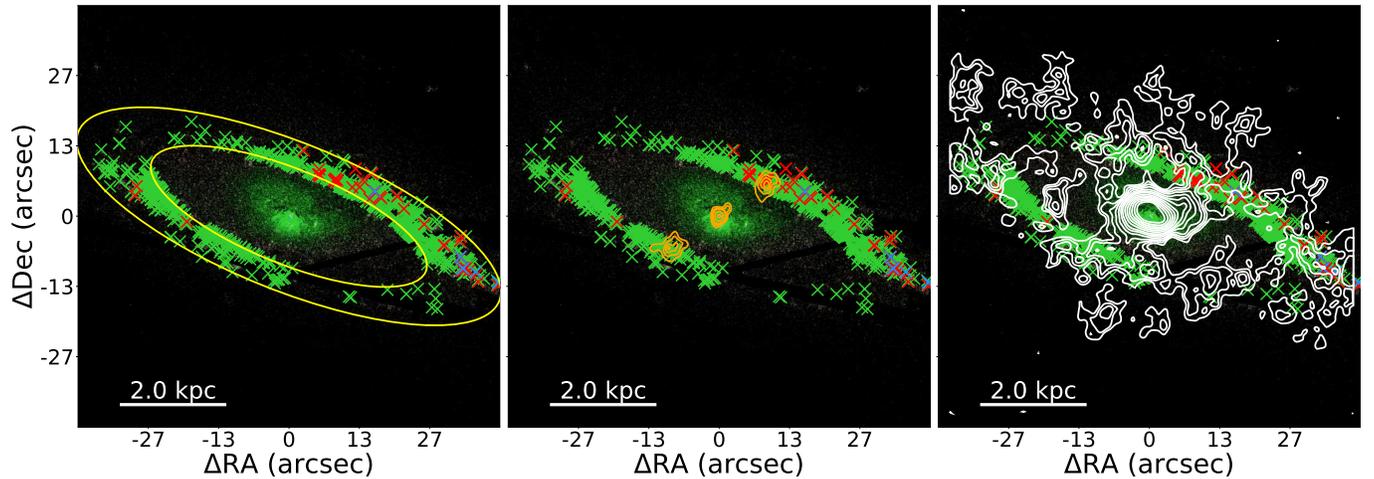

**Figure 6.** Left panel: spatially resolved excitation map of the inner 8 kpc region, corresponding to Map 1 in Figure 4. LINER-like, Seyfert-like, and cocoon-like pixels within the yellow elliptical region are marked with green, red, and blue X's, respectively. Center panel: same as the left panel, overlaid with low-resolution 1.5 GHz eMERLIN radio contours (in orange), obtained using a uv-tapered scale of 200 k$\lambda$ (R. D. Baldi et al. 2018). Right panel: same as the center panel, with CO(1-0) zeroth-moment contours from K. Sakamoto et al. (2000) shown in white. In all panels, north is up and east is to the left.

Additionally, the central CO disk shows a steep velocity gradient, with velocities spanning $\sim 750$ km s$^{-1}$, about 30% higher than in the rest of the galaxy (K. Sakamoto et al. 2000). This kinematic signature implies significant turbulence and shock activity, likely linked to inflows from the outer CO ring into the nuclear CO disk. We therefore conclude that shocks associated with gas inflow are a likely contributor to the large-scale LINER-like emission in this region.

### 4.2.3. Star-forming Ring

The H$\alpha$ narrow-line image in Figure 2 reveals a clumpy, ring-shaped structure of enhanced line emission, with dimensions of $r_{\mathrm{maj}} \sim 4$ kpc and $r_{\mathrm{min}} \sim 1.2$ kpc (Figure 7). In the sr-S-BPT maps (Figure 4, Map 1), this structure is composed predominantly of H II–like pixels (97.18%, see also Section 4.2.2), indicating that its ionization is primarily driven by young stellar populations.

Figure 7 overlays the H$\alpha$ image with the 1.5 GHz eMERLIN radio contours (uv-tapered at 200 k$\lambda$) from R. D. Baldi et al. (2018; shown in blue). Regions of enhanced, clumpy H$\alpha$ emission are spatially coincident with the southern radio lobe's apparent point of impact with the ISM, suggesting a direct, localized interaction between the jet and ionized gas. In this region, the H$\alpha$ flux is $\sim 60\%$ higher than in a comparable portion of the ring, supporting the idea that shocks induced by the jet–ISM interaction may contribute to the observed line enhancement.

Figure 7 also includes CO(1-0) contours from K. Sakamoto et al. (2000; in white). The H$\alpha$ ring is consistent with a bar lane associated with the small stellar bar proposed by K. Sakamoto et al. (2000; see Section 4.2.2), which likely channels gas into both the outer CO ring and the central condensation disk.

## 5. Conclusions

We used narrowband HST imaging to construct S-BPT diagrams and spatially resolved S-BPT (sr-S-BPT) maps of an $80'' \times 80''$ (8 kpc$^2$) region in the NLR of the LINER I AGN NGC 5005. This is the first study of its kind in a LINER-dominated galaxy, allowing us to explore the interplay between different excitation mechanisms with unprecedented spatial detail.

Our analysis reveals that the nucleus of NGC 5005 is dominated by Seyfert-like emission, consistent with AGN photoionization. Chandra/ACIS-S observations confirm the presence of a hard X-ray source colocated with the Seyfert-like nucleus, strengthening the case for AGN-driven excitation in the central region.

Surrounding the Seyfert-like pixels, we identify a previously undetected, thin LINER-like cocoon, likely driven by shock excitation. Thanks to the high spatial resolution of HST, this study is the first to resolve the transition between Seyfert-like, cocoon-like, and LINER-like excitation regions in a galaxy. The LINER-like cocoon has a characteristic thickness of $\sim 10$–20 pc, though deeper observations may reveal additional structures.

LINER-like emission in NGC 5005 is found on two distinct spatial scales:

1. A compact central LINER region ($r \sim 1$ kpc), which encloses the Seyfert-like nucleus and cocoon. This region is likely photoionized by the low-luminosity AGN located at the optical centroid, with additional contributions from shocks.
2. A larger-scale LINER-like zone ($r \gtrsim 2$ kpc), which surrounds the central LINER region and overlaps with the outer star-forming ring. This extended LINER-like emission could be powered by ionization from post-AGB stars, although localized shocks from inflowing gas may also contribute.

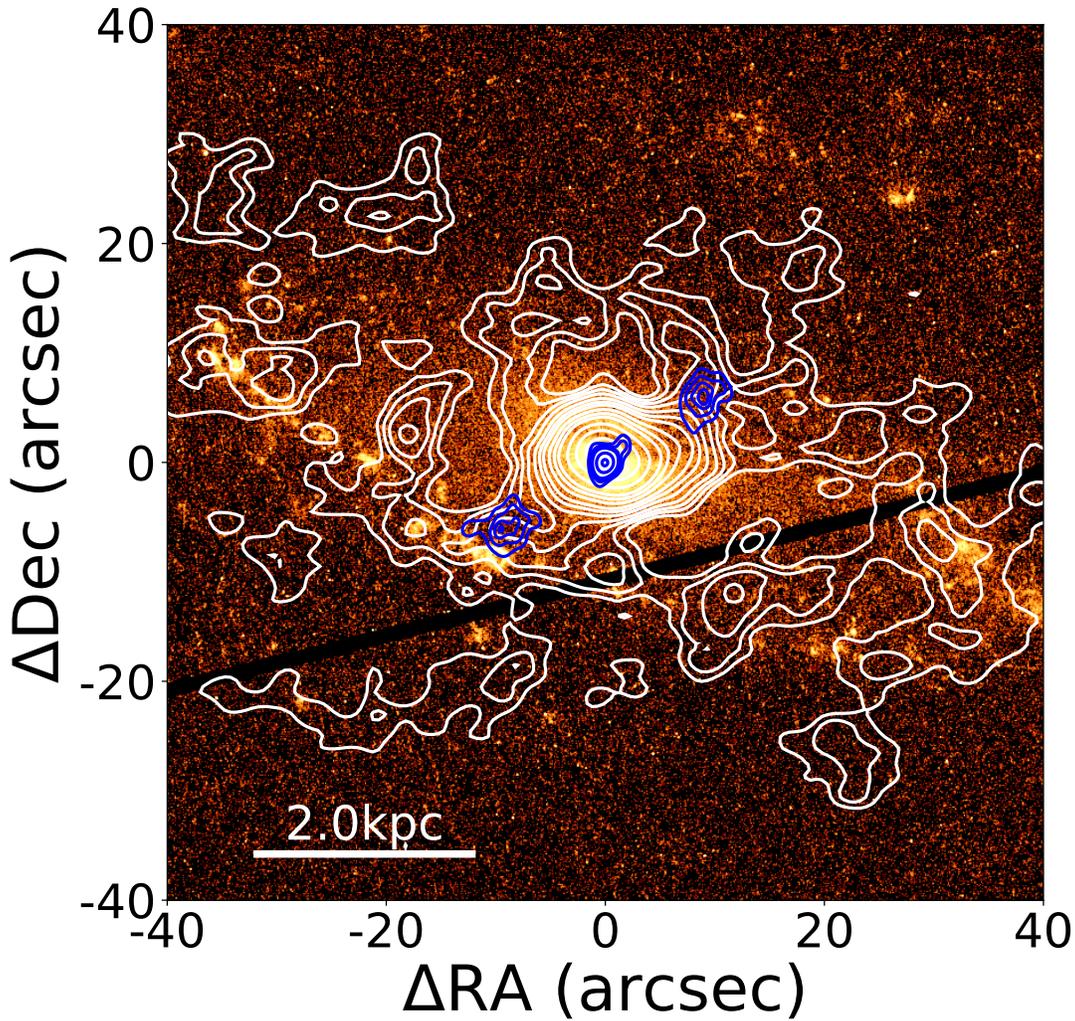

**Figure 7.** Hα narrow-line image of NGC 5005. Overlaid in blue are the low-resolution 1.5 GHz eMERLIN radio contours from R. D. Baldi et al. (2018), and in white are the CO(1-0) zeroth-moment contours from K. Sakamoto et al. (2000). North is up and east is to the left.

H II–like regions are also present at both small and large scales:

1. In the inner 500 pc, the H II–like emission likely results from shocks associated with nuclear radio jet activity, which may induce localized star formation. Alternatively, an energetic outflow could also explain the H II–like bubble observed in this region.
2. At $r \sim 4$ kpc, H II–like emission dominates in a ring-like structure, tracing a large-scale star-forming ring excited by young stellar populations.

This study provides new insights into the ionization structure of LINER-dominated galaxies and highlights the complex role of AGN/stellar feedback, shock excitation, and different stellar populations in shaping their emission-line properties. However, to fully disentangle these mechanisms on a region-by-region basis, particularly where shocks are involved, high-resolution kinematic data from IFU spectroscopy with comparable spatial resolution is essential. A multiwavelength analysis at matched spatial resolution may provide robust diagnostics of the different ionization mechanisms and deepen our understanding of AGN–host galaxy interactions and feedback processes.

**Acknowledgments**

This work was supported by HST grant GO-16837.001-A (PI: Fabbiano), and partially by NASA contract NAS8-03060 (CXC). PyRAF and STSDAS are products of the Space Telescope Science Institute, which is operated by AURA for NASA. We thank R. D. Baldi and K. Sakamoto for kindly providing a portion of the data sets used in this work. We thank the anonymous referee for helpful comments that improved the quality of this paper.

# Appendix
# Reddening Correction

This appendix outlines the methodology used in Section 2.3 to correct the HST images for reddening prior to continuum subtraction. The approach employs a reddening law (D. Calzetti et al. 2000) and calculates extinction on a pixel-by-pixel basis. This ensures that intrinsic fluxes are accurately estimated by accounting for wavelength-dependent extinction effects.

### A.1. Methodology

The observed flux at a given wavelength, $F_{\rm obs}(\lambda)$, relates to the intrinsic flux, $F_{\rm int}(\lambda)$, by

$$F_{\rm obs}(\lambda) = F_{\rm int}(\lambda) \times 10^{-0.4 \cdot A_\lambda}, \quad (A1)$$

where $A_\lambda$ is the extinction at wavelength $\lambda$. Given two continuum bands, namely 1 and 2, the ratio of their observed fluxes can be expressed as

$$\frac{F_{\rm obs}(1)}{F_{\rm obs}(2)} = \frac{F_{\rm int}(1) \times 10^{-0.4 \cdot A_1}}{F_{\rm int}(2) \times 10^{-0.4 \cdot A_2}}. \quad (A2)$$

Assuming that the unabsorbed color ratio $C$ of stars in bands 1 and 2 is constant across a given region, i.e.,

$$C = \frac{F_{\rm obs}(1)}{F_{\rm obs}(2)} = \frac{F_{\rm int}(1)}{F_{\rm int}(2)}, \quad (A3)$$

where $A_1 = A_2 = 0$ (i.e., no reddening), the extinction at a wavelength, $A_\lambda$, may be related to the color excess $E(B-V)$, using the reddening curve $k(\lambda)$:

$$A_\lambda = k(\lambda) \cdot E(B-V) = \frac{k(\lambda) \cdot A_V}{R_V}. \quad (A4)$$

Combining Equations (A2), (A3), and (A4), we derive

$$\frac{F_{\rm obs}(1)}{F_{\rm obs}(2)} = C \cdot 10^{-0.4 \cdot E(B-V) \cdot (k_1 - k_2)}, \quad (A5)$$

which allows $E(B-V)$ to be calculated as

$$E(B-V) = \frac{1}{-0.4 \cdot (k_1 - k_2)} \cdot \log\left(\frac{F_{\rm obs}(1)}{C \cdot F_{\rm obs}(2)}\right). \quad (A6)$$

### A.2. Pixel-by-pixel Extinction Calculation

Using a continuum ratio map (see Figure 1), defined as $c_{\rm map} = \frac{F_{\rm obs}(1)}{F_{\rm obs}(2)}$, $E(B-V)$ can be computed for each pixel as

$$E(B-V) = \frac{1}{-0.4 \cdot (k_1 - k_2)} \cdot \log\left(\frac{c_{\rm map}}{C}\right), \quad (A7)$$

where $k_1$ and $k_2$ are the extinction curve values for the respective continuum filters. These values can be obtained from resources such as the Spanish Virtual Observatory website,[10] which provides $k(\lambda)$ for various HST filters.

The obtained reddening correction is then applied to each continuum filter before subtraction from the narrow-line image. This process essentially converts the continuum image into a specific continuum image for a given narrow band. This methodology allows for correction of reddening effects on a pixel-by-pixel basis, ensuring that the final corrected fluxes reflect intrinsic values as accurately as possible given observational constraints.

## ORCID iDs

Anna Trindade Falcão  https://orcid.org/0000-0001-8112-3464
G. Fabbiano  https://orcid.org/0000-0002-3554-3318
M. Elvis  https://orcid.org/0000-0001-5060-1398
S. Kraemer  https://orcid.org/0000-0003-4073-8977
W. P. Maksym  https://orcid.org/0000-0002-2203-7889
D. L. Król  https://orcid.org/0000-0002-3626-5831

---
[10] http://svo2.cab.inta-csic.es/svo/theory/fps/index.php?mode=browse&gname=HST